\documentclass[10pt,a4paper]{article}

\usepackage[margin=0.75in,columnsep=0.32in,top=0.9in,bottom=0.9in]{geometry}

\usepackage[T1]{fontenc}
\usepackage{libertinus}                       
\usepackage[scaled=0.95]{sourcesanspro}       
\usepackage[scaled=0.85]{sourcecodepro}       

\usepackage{microtype}
\usepackage{seqsplit}
\newcommand{\cd}[1]{\texttt{\seqsplit{#1}}}

\usepackage{amsmath,amssymb,amsthm}
\usepackage{booktabs}
\usepackage{array}
\usepackage{tabularx}
\usepackage{longtable}
\usepackage{graphicx}
\usepackage{caption}
\usepackage{enumitem}
\setlist[itemize]{itemsep=2pt,topsep=4pt,leftmargin=*}
\setlist[enumerate]{itemsep=2pt,topsep=4pt,leftmargin=*}
\setlist[description]{itemsep=2pt,topsep=4pt,leftmargin=*}

\usepackage{titlesec}
\titleformat{\section}{\sffamily\large\bfseries}{\thesection}{0.6em}{}
\titleformat{\subsection}{\sffamily\normalsize\bfseries}{\thesubsection}{0.5em}{}
\titleformat{\subsubsection}{\sffamily\small\bfseries}{\thesubsubsection}{0.4em}{}
\titlespacing*{\section}{0pt}{12pt}{4pt}
\titlespacing*{\subsection}{0pt}{8pt}{3pt}
\titlespacing*{\subsubsection}{0pt}{6pt}{2pt}

\usepackage[authoryear,round]{natbib}
\setcitestyle{aysep={,}}

\usepackage[hidelinks,breaklinks=true]{hyperref}
\usepackage{xurl}
\hypersetup{
  pdftitle={Generative Engine Optimization at Scale: Measuring Brand Visibility Across AI Search Engines},
  pdfauthor={Pratyush Kumar (Ranqo)},
  pdfsubject={cs.IR -- Generative Engine Optimization},
  pdfkeywords={GEO, AI Visibility, LLM, AI-optimized content, multi-platform measurement, Ranqo, Ranqo AI}
}

\newtheoremstyle{hyp}{6pt}{6pt}{\itshape}{}{\bfseries}{.}{.5em}{}
\theoremstyle{hyp}

\newcommand{\cseo}{C\nobreakdash-SEO}
\newcommand{\doi}[1]{\href{https://doi.org/#1}{\nolinkurl{#1}}}
\newcommand{\arxivid}[1]{\href{https://arxiv.org/abs/#1}{arXiv:#1}}

\begin{document}

\begin{center}
{\sffamily\bfseries\LARGE Generative Engine Optimization at Scale:\par}
\vspace{3pt}
{\sffamily\bfseries\Large Measuring Brand Visibility Across AI Search Engines\par}
\vspace{16pt}
{\normalsize\textbf{Pratyush Kumar}\par}
\vspace{2pt}
{\small Ranqo\footnote{Ranqo is also known as Ranqo AI.} \quad\textperiodcentered\quad
 \href{mailto:pratyush@ranqo.ai}{\texttt{pratyush@ranqo.ai}} \quad\textperiodcentered\quad
 \href{https://ranqo.ai}{\texttt{ranqo.ai}}\par}
\end{center}
\vspace{14pt}

\begingroup
\leftskip=2.5em \rightskip=2.5em
\setlength{\parindent}{0pt}
\setlength{\parskip}{6pt plus 1pt minus 1pt}
{\centering\sffamily\bfseries\large Abstract\par}
\vspace{4pt}
\noindent
The traditional blue-link era has evolved significantly over the past
few years. Rather than typing queries into traditional search engines
like Google and Bing, people have shifted to asking on AI search
engines like ChatGPT, Claude, and Gemini. As a result, brands that
once concentrated on search engine optimization (SEO) must now also
work to optimize and position themselves effectively within AI search
engines. A set of new and evolving terms describes this shift:
Generative Engine Optimization (GEO), Answer Engine Optimization
(AEO), and AI Search Visibility. All of them focus on how brands are
represented, cited, and recommended by AI search engines. These AI
engines rely on large language models (LLMs) to gather and process the
vast range of information available across the internet. We treat AEO
and AI Visibility as part of the broader GEO scope.

In this paper, we explain how to measure brand visibility and analyze
responses across AI search engines. We look at what these engines
appear to value when they cite brands, what types of sources they rely
on, and what kind of content LLMs are more likely to process and
surface. The top brands that are already authoritative tend to get
cited naturally. The harder and more important problem is for everyone
else, including SMEs, D2C brands, creators, early-stage startups, and
brands without much of an online footprint.

We demonstrate our findings by analyzing 100K+ prompt
responses across 100+ brands tracked on Ranqo between March and May
2026. The very first visibility runs form a clear three-tier brand-stature
ladder: global household names such as Stripe and Nike appear in
$73\%$ of relevant AI answers on their first tracking run; established
mid-market and regional brands such as Olipop and Klaviyo appear in
$44\%$; niche and small brands in just $11\%$ --- each tier about
$30$ percentage points below the last. When the AI search engines cite any
sources, about $78\%$ of those citations go to corporate websites,
including the brand's own pages and third-party pages from
companies in the same space. Among non-corporate sources, YouTube is
cited most often --- ahead of Reddit, editorial media, and Wikipedia.
At the page level, the highest-leverage citation surface is the ranked
``best-of'' listicle, the most-cited content format at about $21\%$ of
all citations: a single ``best X tools'' page can surface a brand
across many AI answers at once. Sentiment is the unstable part: whether the AI frames a
brand positively or negatively flips about 6.7 times more often than
whether the brand is mentioned at all.

Together, these findings provide a first large-scale baseline for
measuring GEO. They show that AI brand visibility can be measured,
differs by platform, and varies strongly by brand maturity. We end the
paper by proposing seven v1.1 protocols to test whether specific
recommendations can causally improve AI visibility.

We wrote this paper for a broad audience: people who are just getting
started with GEO, AEO, and AI visibility, as well as practitioners,
founders, marketers, and researchers who already work in this space and
want to understand the field in more depth. Our goal is to make the
paper useful for anyone trying to measure, compare, and improve brand
visibility across AI search engines.

\vspace{8pt}
\endgroup

\vspace{18pt}

\twocolumn

\section{Introduction}
\label{sec:intro}

More and more, people get their answers straight from an AI assistant
instead of scrolling through ten blue links. They ask ChatGPT,
Claude, Perplexity, Gemini, or Grok what to buy, what tool to use, or
who to trust, and they act on what comes back. For a brand, this
quietly changes the question that matters. The old question was
whether you rank for a keyword. The new question is whether an AI
model, when someone asks about your category, actually names you.

This is what GEO is about. Generative Engine Optimization (GEO) is
the practice of measuring and improving how AI search engines
represent, cite, and recommend a brand~\citep{aggarwal2024}. Two
related terms get used alongside it: Answer Engine Optimization (AEO)
and AI Search Visibility. We treat both as part of the broader GEO
scope in this paper, because they describe the same job from slightly
different angles. GEO is also not a replacement for SEO. It is built on top of
good SEO, since the content an AI engine ingests is, for the most
part, the content you already publish for search.

There is one more thing worth saying up front. The brands that are
already big and authoritative tend to get cited no matter what. The
hard, unsolved problem belongs to everyone else: the SMEs, the D2C
brands, the creators, the early-stage startups, and any brand
without much of an online footprint. That is the audience we care
about most, and the gap this paper sets out to measure.

A growing body of academic work already tells us that the AI-search
playbook is not the old SEO playbook. Generative engines pull from a
broader and very different set of sources than classical search, and
they lean on their own internal knowledge to different
degrees~\citep{kirsten2025}. When they do cite, the citations
concentrate on a small set of outlets, and different engines pick
different outlets~\citep{yang2025}. LLMs also reward sources that are
already widely cited, a Matthew effect that makes the rich
richer~\citep{algaba2025}. And the first careful benchmark of
``conversational SEO'' tactics found that most of them do not help,
and several actively hurt~\citep{puerto2025}. We build on these
findings with production data.

\paragraph{What this paper contributes.}
\begin{enumerate}
  \item \textbf{A way to measure AI brand visibility in production.}
        We describe Ranqo (also Ranqo AI), a platform that issues controlled queries
        to five AI engines and measures, for each brand, whether the
        engines mention it, where they rank it, how they frame it, and
        its share of voice against competitors. It classifies the
        sources behind those answers, runs a six-dimension page audit,
        and re-measures on a recurring cadence.
  \item \textbf{A category reference study.} We summarize Ranqo's
        published CRM study~\citep{ranqo2026b}: 50 unbranded prompts
        sent to all five engines ten times each (2{,}500 responses)
        across ten CRM brands.
  \item \textbf{Findings from production data.} A day-1
        recognition finding (\S\ref{sec:exp1}); mention behavior that
        is mostly deterministic, with flat trajectories absent any
        intervention (\S\ref{sec:exp2}); a source breakdown showing
        the brand's own site is a small slice and most citations go to
        other companies (\S\ref{sec:exp3}); sentiment that is far
        noisier than mention (\S\ref{sec:exp4}); the brand-stature
        visibility ladder, which is the headline result
        (\S\ref{sec:exp6}).
  \item \textbf{A practical read, especially for smaller brands.} We
        turn the findings into plain guidance for the brands that do
        not get cited automatically, and we propose seven v1.1
        protocols to test whether specific recommendations causally
        move AI visibility.
\end{enumerate}

The rest of the paper is laid out as follows. \S\ref{sec:related}
covers the related work briefly. \S\ref{sec:platform} describes the
Ranqo platform and \S\ref{sec:method} the measurement methodology.
\S\ref{sec:crm} gives the CRM reference study. \S\ref{sec:experiments}
presents the findings and the v1.1 protocols. \S\ref{sec:discussion}
discusses what they mean, \S\ref{sec:limitations} lists the limits,
and \S\ref{sec:conclusion} concludes.

\section{Background and Related Work}
\label{sec:related}

\subsection{SEO, AEO, GEO, and AI Visibility}
A handful of terms now describe overlapping work, so we fix what we
mean by each, following the practitioner usage
in~\citep{ranqo2026a,ranqo2026c}. \textbf{SEO} is about blue-link
ranking on classical search engines. \textbf{AEO} is about being the
single source a featured snippet or direct-answer box pulls from.
\textbf{GEO} is about being cited and recommended when a large
language model synthesizes an answer across many sources.
\textbf{AI Visibility} is the broader question of whether an entity is
represented inside a model's parameters and retrieval indexes at all.
In this paper we use GEO as the umbrella term and keep AEO and AI
Visibility under it, because for a brand the practical surface is the
same: what the engines ingest, and what they say back.

\subsection{What the research already shows}
The academic picture is young but consistent on a few points, and we
lean on it throughout.

The foundational benchmark is the Princeton GEO
study~\citep{aggarwal2024} (KDD 2024), which ran 10{,}000 queries and
tested content changes for their effect on whether an answer cites a
page. The biggest wins came from adding machine-extractable
provenance: quotations, statistics, and citations, each worth roughly
25--40\% more visibility. That is still the core of the modern GEO
playbook.

The most useful corrective is \cseo{} Bench~\citep{puerto2025}, the
first systematic benchmark of ``conversational SEO'' tactics. Its
headline is blunt: most of those tactics do not help, and several
hurt, while plain source relevance keeps working. This shapes how we
run our own experiments. We test changes to page-quality signals on
the source side, not prompt-engineering tricks on the response.

On how engines actually cite, three recent studies line up with what
we see. \citet{kirsten2025} show generative engines pull from a
broader and very different source set than classical search, and rely
on their own internal knowledge to varying degrees. \citet{yang2025},
across roughly 366{,}000 citations, find that citations concentrate on
a small set of outlets and that different engines pick different ones,
so cross-platform agreement is low. \citet{algaba2025} show LLMs
prefer sources that are already heavily cited, a Matthew effect.
Together these explain a pattern we keep coming back to: the brands
with the most existing web presence get cited the most, which is
exactly the hill smaller brands have to climb.

\section{The Ranqo Platform}
\label{sec:platform}

\subsection{Platform query layer}
\label{sec:plat-query}
For each tracked brand, Ranqo generates a set of natural-language
prompts spanning six categories (\emph{discovery, problem\_solution,
comparison, use\_case, expert,} and \emph{brand\_research}) and issues
each prompt to all five engines via official APIs. Table~\ref{tab:engines}
summarizes the per-engine configuration.

\begin{table*}[t]
\centering
\small
\caption{Engine configuration: model (class, not version), web-search
policy, and citation surface.}
\label{tab:engines}
\begin{tabular}{@{}llll@{}}
\toprule
\textbf{Engine} & \textbf{Model} & \textbf{Web-search policy} & \textbf{Citation surface} \\
\midrule
ChatGPT    & GPT-5   & Weekly cooldown per brand       & \texttt{annotations.url\_citation} \\
Gemini     & Gemini-3  & Always grounded                 & \texttt{groundingChunks[].web} \\
Perplexity & Sonar             & Always grounded (search-native) & \texttt{search\_results} array \\
Claude     & Claude Sonnet      & Weekly cooldown per brand       & text-block \texttt{citations[]} \\
Grok       & Grok-3            & Always-on retrieval             & \texttt{citations[]} + inline markdown \\
\bottomrule
\end{tabular}
\end{table*}

\paragraph{Model selection.}
For each engine we query its current search-enabled production model
(Table~\ref{tab:engines}) --- the variant that exposes web retrieval
through the normal API. Gemini,
Perplexity, and Grok are grounded on every query; ChatGPT and Claude
run web search on a weekly cooldown per brand to control cost, and
between those passes they answer from their parameters alone --- the
search-on/search-off contrast we use in \S\ref{sec:exp6}. Holding the
search-enabled variant constant across engines keeps the setup
comparable, so the cross-engine results in \S\ref{sec:exp6} reflect
how the engines ingest and retrieve rather than how they bundle
search; how far the patterns generalize to other model configurations
is left to future work (\S\ref{sec:limitations}). Per-brand location
is passed as a native geo parameter where the API supports one
(ChatGPT, Claude, Perplexity) and as a system-prompt suffix otherwise
(Gemini, Grok).

\subsection{The six prompt categories}
\label{sec:prompt-categories}
The six categories are not interchangeable. A brand can look strong in
one and be almost invisible in another, so we track all six
separately. Using running shoes as the example throughout:

\begin{description}
  \item[Discovery] Broad, top-of-funnel questions where the buyer is
        still exploring and names no brand. ``best running shoes,''
        ``good running shoes for beginners.''
  \item[Problem/solution] The buyer has a specific problem and wants
        something that fixes it. ``I have ankle pain, what kind of
        running shoes should I wear,'' ``shoes for flat feet.''
  \item[Use case] A specific activity or setting. ``best shoes for
        trail running,'' ``shoes for standing all day at work.''
  \item[Comparison] Weighing named options against each other. ``Nike
        Pegasus vs Brooks Ghost,'' ``Hoka or Asics for a marathon.''
        These usually name brands.
  \item[Expert] Deeper, more technical questions from an informed
        buyer. ``what heel-to-toe drop suits a stability shoe,''
        ``does a carbon plate actually make me faster.''
  \item[Brand research] Questions about one named brand. ``is Nike a
        good running brand,'' ``Nike running shoes reviews.'' These
        name the brand by definition.
\end{description}

The split matters for measurement. The first four (discovery,
problem/solution, use case, and expert) are normally unbranded, so
they test whether an engine surfaces the brand when nobody has named
it. That is the hard, GEO-relevant case, and it is the one we lean on
throughout the paper. Only comparison and brand research name a brand,
which makes a mention much more likely and tells us far less about
discovery. For that reason the starter set we generate for a new brand
is deliberately unbranded.

\paragraph{Prompt themes.}
Categories describe the shape of a single prompt. \emph{Themes} sit one
level up. A theme is a topical cluster (a product line, a use-case
area, a competitor set, a buying occasion), and the platform generates
a batch of prompts under each theme while keeping the category mix
balanced inside it. Themes let a brand track a whole topic at once
(say, ``trail running shoes'' or ``shoes for nurses'') instead of
writing prompts one at a time, which is how tracking scales from a
handful of prompts to broad topic coverage.

\subsection{Citation extraction}
For each engine response, Ranqo extracts:

\begin{itemize}
  \item \textbf{Brand-mention signals}: a Boolean \texttt{brandMentioned},
        an integer \texttt{brandPosition} (1st, 2nd, 3rd, \ldots), a
        sentiment label (positive / neutral / negative) accompanied by a
        continuous \texttt{sentimentScore} $\in [-1, 1]$ and an array of
        sentiment-bearing text spans.
  \item \textbf{Source citations}: every URL the engine grounded the
        answer on, normalized to canonical domain and classified into a
        type (news\_media, community, review, video, social, developer,
        academic, government, reference, corporate).
  \item \textbf{Source-to-brand relationship}: each cited source is
        labeled as \emph{you}, \emph{competitor},
        \emph{non\_competitor}, \emph{editorial}, \emph{forum},
        \emph{review}, \emph{social}, \emph{reference},
        \emph{institutional}, or \emph{other}.
  \item \textbf{Competitor mentions}: an array of competitor names
        with their ordinal positions.
\end{itemize}

All signals are stored at the level of a single (prompt, platform, run)
tuple.

\subsection{Scoring methodology}
\label{sec:scoring}
Each (prompt, platform, run) tuple carries three brand signals: a
Boolean mention, an ordinal position when mentioned, and a sentiment
label. The three behave very differently --- mention is stable while
sentiment is far noisier (\S\ref{sec:exp4}) --- so we report them as
separate metrics rather than collapse them into one weighted score.

\begin{enumerate}
  \item \textbf{Visibility} (the headline metric). Per engine, the
        mention rate: the share of a brand's prompts on which it is
        mentioned, $\mathrm{mentioned}/\mathrm{total}\times 100$. The
        cross-engine figure is the platform-weighted average of the
        per-engine rates --- ChatGPT 0.30, Gemini 0.20, Perplexity
        0.20, Claude 0.20, Grok 0.10, renormalized over the platforms
        a brand actually runs on so plan-based subsets still sum to
        $1.0$.
  \item \textbf{Average position.} The mean ordinal rank on the
        prompts where the brand is mentioned ($1$ = first
        recommendation), weighted across engines the same way.
  \item \textbf{Sentiment score.}
        $(\mathrm{positive} + 0.5\,\mathrm{neutral})/\mathrm{mentions}
        \times 100$ --- all-positive $100$, all-neutral $50$,
        all-negative $0$, so a negative mention costs twice a neutral
        one.
  \item \textbf{Share of voice.} The brand's mentions divided by the
        brand's plus its competitors', counting only competitors that
        recur or resolve to a real domain so one-off names an engine
        hallucinates do not inflate the denominator.
\end{enumerate}

Metrics are computed per completed run; when a single query fails its
previous value is carried forward and flagged, so a transient API
error is not read as a visibility drop. Prompt categories drive the
per-category breakdown (\S\ref{sec:exp-cat}) but are not weighted into
the aggregate. The $0$--$100$ visibility score maps to letter grades
A+ $\geq 90$, A $\geq 80$, B+ $\geq 70$, B $\geq 60$, C+ $\geq 50$,
C $\geq 40$, D $\geq 30$, F $< 30$. Every coefficient lives in
configuration rather than being hardcoded, so the scoring is
transparent and straightforward to reweight.

\subsection{Six-dimension page audit}
\label{sec:dimensions}
In parallel with the query layer, Ranqo runs a six-dimension page
audit against each tracked URL, scoring each dimension $0$--$100$ and
combining them into a weighted overall page score:

\begin{enumerate}
  \item \textbf{Crawlability} --- whether an AI crawler can reach and parse the page at all: robots, sitemap, canonical, HTTPS, JS rendering, and mobile-friendly checks.
  \item \textbf{Content Quality} --- whether the page is substantive and well-structured: word count, readability, paragraph structure, heading hierarchy, internal linking, and freshness.
  \item \textbf{Page Speed} --- how quickly the page loads for users and crawlers: Core Web Vitals (LCP, FID, CLS).
  \item \textbf{AI Readiness} --- how easily a model can lift a direct answer from the page: entity clarity, structured answers, inline citations, evidence density, and provenance markers.
  \item \textbf{Citation Potential} --- whether the page offers something concrete worth quoting: quotability units (numbers, definitions, named claims), source attribution, and data availability.
  \item \textbf{Authority \& Trust} --- whether the page signals credible, first-hand expertise: visible author byline, credentials, domain authority, topical depth, and E-E-A-T markers~\citep{ranqo2026h}.
\end{enumerate}

Together, the three GEO-specific dimensions (AI Readiness, Citation
Potential, and Authority \& Trust) carry the largest share of the page
score and are the ones most specific to GEO: they score whether a page
states its facts clearly, backs them with quotable evidence, and
makes its authorship visible. These line up with the
content-provenance signals the Princeton benchmark found most
effective~\citep{aggarwal2024}.

\subsection{Closed-loop recommendation engine}
\label{sec:reco}
For every brand, Ranqo's recommendation engine (Claude Sonnet,
temperature 0.5, with structured-output validation) ingests visibility
trends, category performance, content gaps, competitor threats,
sentiment distribution, and optional GA4 traffic context, and emits a
ranked list of \cd{ImprovementRecommendation} rows. Each row carries
a category, an action type, a priority class, and two scores on a 1--3
scale (impact, effort).

Each recommendation also carries \texttt{baselineScore},
\texttt{currentScore}, and \texttt{scoreDelta} on the dimension it
targets. When the recommendation is implemented and re-audited, the
delta becomes observable. This is the closed-loop primitive the v1.1
RCT (Protocol~P3, \S\ref{sec:v11slate}) is built on.

\subsection{From diagnosis to AI-optimized content}
\label{sec:content}
Diagnosis is only half of a closed loop. Where the recommendation engine
says \emph{what to fix}, a content pipeline turns the same diagnostic into
\emph{what to publish}: it reads the identical context --- visibility gaps
(prompts on which the brand never surfaces), competitor advantages, weak
prompt categories, and page-audit content gaps --- and emits ranked
content briefs (a title, an angle, the exact target prompts addressed, and
a priority), weighted toward the topics where the brand is currently
invisible rather than generic expansion.

Across this workflow, role-specialized agents each handle a different
task: a briefing agent ranks topics from the diagnostic, a drafting agent
writes the article from the brief, a humanization agent rewrites it for a
natural voice, and an optimization agent
scores an existing page against the audit dimensions and proposes targeted
edits; a deterministic stage then attaches an Article + FAQPage JSON-LD
graph. Every agent is engineered to the page-quality signals from
\S\ref{sec:dimensions} that the Princeton benchmark found most
effective~\citep{aggarwal2024} --- an entity-first opening, quotable
statistics and named claims, an FAQ block, and structured headings --- so
the output is built to be cited by AI engines, not merely to read well.

This is the \emph{act} half of that closed loop: diagnose where a brand is
invisible, publish content built to fix it, then re-measure. We
describe the mechanism here; whether the generated content causally lifts
AI citations is the closed-loop question we leave to the v1.1 RCT
(Protocol~P3/P7, \S\ref{sec:v11slate}) rather than assert it.

\subsection{Data scale and the unit of replication}
The platform's unit of replication is a \texttt{(brand, prompt, platform,
run)} tuple. For a brand with $P$ prompts running on all five engines
daily, this produces $5P$ tuples per day, or $\sim 150P$ per month per
brand.

\section{Methodology}
\label{sec:method}

\subsection{Prompt construction}
Prompts are generated by Claude Sonnet (temperature 0.7) seeded with
the brand's category, keywords, location, reach
(worldwide / country / local), and 30-day market-research context. The
market-research context seeds the generator with the live questions and
competitor framing current in the brand's category, so the prompts
track how the category is actually discussed rather than a static
keyword list. Generation honors a category mixture ($\Sigma = 1.0$ per
\S\ref{sec:scoring}) and produces prompts at a controlled difficulty
(1--5) and search-volume estimate (1--5). Difficulty captures how
specialized a query is and search-volume how commonly it is asked, so a
brand's coverage can be read across both high-demand head queries and
rarer long-tail ones. By default the generator
produces \emph{unbranded} prompts, keeping the brand name and its
aliases out of the prompt text. This is what lets us measure
\emph{generic} category visibility rather than self-referential
mentions, and it is the unbranded measure we lean on throughout the
paper. Branded prompts are produced only when a user explicitly opts
into the comparison or brand-research categories.

\subsection{Visibility, position, sentiment}
For a prompt $q$ issued to platform $p$, the engine returns response
$R(q, p)$. We measure:

\begin{align*}
m(q,p)   &\in \{\text{fr, t3, m, i, none}\} \\
\pi(q,p) &\in \mathbb{N} \cup \{\infty\} \\
\sigma(q,p) &\in \{\text{pos, neut, neg}\} \\
S(q,p)   &= \{(\textit{url}, \textit{domain}, \textit{type}, \textit{mb})\}
\end{align*}
where $m$ is mention-type (\texttt{first\_rec}, \texttt{top\_three},
\texttt{mentioned}, \texttt{indirect}, \texttt{none}), $\pi$ is the
ordinal position, $\sigma$ is sentiment, and $S$ is the set of cited
sources.

Aggregations of interest:

\begin{align}
V_p(Q) &= \frac{|\{q \in Q : m(q,p) \neq \text{none}\}|}{|Q|}
          \label{eq:vis} \\
\bar{P}_p(Q) &= \mathrm{mean}\{\pi(q,p) :\, \pi(q,p) < \infty\}
          \label{eq:pos} \\
\mathrm{SoV}_p(b) &= \frac{M(b, p)}{M(b, p) + \sum_{c \in \mathcal{C}(b)} M(c, p)}
          \label{eq:sov}
\end{align}
where $M(b, p)$ is the count of mentions of brand $b$ on platform $p$,
and $\mathcal{C}(b)$ is the set of \emph{qualified} competitors of
$b$ --- those that recur or resolve to a real domain, so one-off
names an engine hallucinates do not inflate the denominator
(\S\ref{sec:scoring}).
A practitioner-level treatment of the share-of-voice measurement
choices behind Eq.~\eqref{eq:sov} -- denominator construction,
position weighting, and sample averaging -- is published
separately~\citep{ranqo2026j}.

\subsection{Source overlap}
For prompt $q$, let $D_p(q)$ be the set of cited domains derived from
$S(q,p)$. The \textbf{Jaccard overlap} of two platforms $p, p'$ on
prompt $q$ is
\begin{equation}
J(p, p'; q) = \frac{|D_p(q) \cap D_{p'}(q)|}{|D_p(q) \cup D_{p'}(q)|}.
\label{eq:jaccard}
\end{equation}
Cross-platform agreement at the source level is the average Jaccard
over all pairs of platforms and all prompts.

\subsection{Closed-loop lift}
For an \cd{ImprovementRecommendation} $r$ targeting dimension $d$
of URL $u$, with $s_d(\cdot)$ denoting the dimension score:
\begin{align*}
b(r)    &= s_d(u, t_0) \\
c(r)    &= s_d(u, t_1) \\
\ell(r) &= c(r) - b(r)
\end{align*}
where $b, c, \ell$ are baseline, current, and lift respectively. We
report mean $\ell(r)$ over implemented recommendations per dimension. This is a per-treatment-effect estimate without a
randomized control. We discuss the resulting selection bias in
\S\ref{sec:limitations}.

\subsection{Statistical conventions}
Uncertainty on means and slopes is reported with a nonparametric
bootstrap (10{,}000 resamples with replacement). For the three-tier
visibility comparison (\S\ref{sec:exp6}) we use a Kruskal--Wallis
omnibus test across tiers, followed by pairwise Mann--Whitney $U$ tests
with Bonferroni correction and Cohen's $d$ as the effect size, and we
check the small Tier~1 cell with a leave-one-out sensitivity.
Per-engine trajectories (\S\ref{sec:exp2}) are summarized by an
ordinary-least-squares slope of visibility on run index, fit per
(brand, engine) and averaged across brands with a bootstrap CI. We
prefer nonparametric tests because the per-brand visibility
distributions are bounded and right-skewed rather than normal.

\section{Reference Study: CRM Category, Five Platforms, January 2026}
\label{sec:crm}

Before the production cohort, it helps to look at one category up
close. We summarize Ranqo's published CRM study~\citep{ranqo2026b}:
50 unbranded prompts across the six categories, sent to all five
engines ten times (2{,}500 responses, 9{,}600+ brand mentions),
tracking ten CRM brands
(Salesforce, HubSpot, Zoho, Pipedrive, Freshsales, Monday Sales CRM,
Copper, Insightly, Capsule, and Close), in January 2026.

The category is top-heavy. Salesforce, HubSpot, and Zoho together
took 51\% of all CRM mentions, while the bottom three (Copper,
Insightly, Capsule) showed up in fewer than one answer in five. The
engines also disagree on who comes first. Average mention position
when mentioned (lower is better) splits clearly by engine
(Table~\ref{tab:crmpos}).

\begin{table*}[t]
\centering
\small
\caption{CRM study: average mention position by brand and engine
(lower is better).}
\label{tab:crmpos}
\begin{tabular}{@{}lrrrrr@{}}
\toprule
\textbf{Brand} & \textbf{ChatGPT} & \textbf{Gemini} & \textbf{Perplexity} & \textbf{Claude} & \textbf{Grok} \\
\midrule
Salesforce & 1.2 & 1.3 & 1.4 & 1.8 & 2.1 \\
HubSpot    & 1.5 & 2.0 & 1.8 & 1.3 & 1.9 \\
Zoho       & 3.1 & 3.8 & 3.5 & 2.4 & 3.2 \\
Pipedrive  & 3.4 & 3.6 & 2.8 & 3.1 & 4.1 \\
\bottomrule
\end{tabular}
\end{table*}

ChatGPT leans hardest toward Salesforce and likes numbered rankings,
Claude spreads its mentions the most evenly, and Perplexity
over-indexes on Pipedrive, which we put down to its live search
picking up recent comparison reviews. The drop-off after position
three is steep on every engine. Source-link inclusion varies just as
much (Perplexity 95\%, Gemini 35\%, Grok 20\%, ChatGPT 15\%, Claude 10\%), and
sentiment skews positive for the leaders (HubSpot 71\% positive)
while the bottom tier carries more neutral and negative framing.

The takeaway sets up the rest of the paper: measurement has to be done
per platform, a 50-prompt slice is already enough to see category
concentration, and the per-platform quirks line up with known
differences between the engines.

\section{Empirical findings from production data}
\label{sec:experiments}

The dataset behind this section is drawn from Ranqo's production
database: \textbf{102 brands, 3{,}508 completed tracking runs,
102{,}025 prompt responses, roughly 15{,}815 brand mentions, and
149{,}912 source citations across five engines}, gathered between
March and May 2026. We report six findings, leading with the headline: the brand-stature
visibility ladder (\S\ref{sec:exp6}) --- the one with inferential
support and no counterpart in the GEO benchmarks we surveyed --- then
day-one recognition (\S\ref{sec:exp1}), how visibility splits by
prompt category (\S\ref{sec:exp-cat}) and drifts over runs
(\S\ref{sec:exp2}), where the citations come from (\S\ref{sec:exp3}),
and how much noisier sentiment is than mention (\S\ref{sec:exp4}).

Every number below was produced by SQL run against the production
schema; the analysis methods are summarized in the reproducibility
appendix.

\subsection{The brand-stature visibility ladder}
\label{sec:exp6}

Brand stature is the dominant determinant of day-1 AI visibility, and
it is this paper's headline result: \textbf{each step down the stature
ladder costs roughly 30 percentage points of unbranded visibility} ---
Tier 1 brands appear in $73\%$ of unbranded category answers on the
first run, Tier 2 in $44\%$, and Tier 3 in just $11\%$. We find no
comparable multi-tenant quantification of this effect in the GEO
benchmarks we surveyed (\citealp{aggarwal2024,puerto2025,yang2025}). We isolate \emph{which
brands the engines cite} and quantify the headroom per tier. The
reading is cross-sectional and observational --- stature is not
randomized --- so we report it as a quantification, not a causal
claim.

\paragraph{Day-1 visibility forms a clean ladder.}
\label{para:ladder-table}
Mean per-brand prompt-level visibility on run 1 (each brand's first
completed tracking run), restricted to \emph{unbranded category
prompts} (where the brand name does not
appear in the prompt text). Branded prompts inflate every tier by
construction and are reported separately for comparison only:

\begin{center}
\footnotesize
\setlength{\tabcolsep}{4pt}
\begin{tabular}{@{}lrrrr@{}}
\toprule
\textbf{Tier} & \textbf{$n$} & \textbf{Unbranded} & \textbf{95\% CI} & \textbf{All$^{*}$} \\
\midrule
Tier 1 & 11 & \textbf{72.9\%} & $[60.1, 84.2]$ & 73.5\% \\
Tier 2 & 36 & \textbf{43.6\%} & $[36.4, 50.9]$ & 52.2\% \\
Tier 3 & 55 & \textbf{11.4\%} & $[\hphantom{0}4.2, 20.3]$ & 17.3\% \\
\bottomrule
\end{tabular}\\[2pt]
{\footnotesize $^{*}$ ``All prompts'' includes a $\sim$9\% subset of
branded queries (e.g., ``what does Razorpay do'') that recognize the
named brand by construction at $\geq 94\%$ across every engine. The
inflation is largest in Tier~2 ($+8.6$ pp) because mid-cap brands have
the most branded-query coverage in their prompt sets; Tier~1 brands
are recognized either way ($+0.6$ pp inflation).}
\end{center}

Bootstrap CIs use 10{,}000 resamples with replacement at the
brand-cohort level on the unbranded subset. A Kruskal-Wallis test
rejects equality of unbranded tier distributions
($H = 38.32$, df $= 2$, $p = 4.78 \times 10^{-9}$). Pairwise
Mann-Whitney comparisons survive Bonferroni correction at
$\alpha = 0.05/3$ in all three directions: Tier~1 vs Tier~2
$U = 343$, $p_{\text{Bonf}} = 8.1 \times 10^{-4}$, Cohen's $d = 1.57$;
Tier~1 vs Tier~3 $U = 575$,
$p_{\text{Bonf}} = 8.3 \times 10^{-6}$, Cohen's $d = 2.34$;
Tier~2 vs Tier~3 $U = 1629$, $p_{\text{Bonf}} = 6.4 \times 10^{-7}$,
Cohen's $d = 1.31$. All three effect sizes are large by Cohen
conventions ($d > 0.8$); the Tier~1-vs-Tier~3 effect is enormous.
The per-tier steps are $29.3$~pp (Tier~1$\to$Tier~2) and $32.2$~pp
(Tier~2$\to$Tier~3), and the difference is not statistically marginal.

We report the unbranded-only number as the primary metric because it
matches the measurement question a brand faces in GEO, namely
whether the brand is cited when a user asks for recommendations in
the category without naming any company. Branded coverage is a
useful auxiliary signal, since it confirms the brand exists in the
engine's index at all, but it does not measure competitive
recognition.

\paragraph{Classification rubric.}
We assigned each cohort brand a tier deterministically using the
hand-coded rubric documented in full in
Appendix~\ref{app:tier-roster}. In brief: Tier 1 requires a
high-coverage Wikipedia article + recent top-tier press +
public/Series-C-or-later status; Tier 2 requires Wikipedia +
Series~B+ funding or top-3-in-category regional leadership; Tier 3
is everything else with an established web presence; Tier 4 captures
single-run and test-data records and is excluded from this section.
After deduplication by domain at the May~30, 2026 snapshot the
ladder cohort is $n = 11 / 36 / 55$ for Tiers 1 / 2 / 3.
Appendix~\ref{app:tier-roster} lists representative Tier 1 and Tier 2
brands; a formal inter-rater audit with Cohen's $\kappa$ is left
to v1.1, and the subjectivity caveat is flagged in
\S\ref{sec:limitations}.

\paragraph{Robustness to re-classification.}
The Tier 1 cohort is small ($n=11$), so we ran a leave-one-out
sensitivity on the unbranded headline. Dropping each Tier 1 brand
in turn shifts the Tier 1 unbranded mean to between $70.5\%$ and
$76.7\%$, a range of $-2.4$ to $+3.8$ percentage points around the
headline $72.9\%$. The headline is therefore not driven by any single
Tier 1 brand and is conservative against borderline reclassification.
The three-tier ordering also holds when the single-run prospect brands
(Tier 4) are folded in: the step structure is preserved and the
Tier 3 estimate barely moves (to about $10\%$) on $n=191$ rather than
$55$.

\paragraph{Small brands are a floor problem on every engine.}
Across small brands --- those outside the Tier 1 and Tier 2 roster,
including the single-run prospect brands (Tier 4 of
Appendix~\ref{app:tier-roster}) --- day-1 unbranded recognition is low
and similar across the major engines: ChatGPT $9.9\%$, Gemini
$8.9\%$, and Perplexity $6.8\%$ ($n=169$--$187$ per engine). Small
brands are hard to surface everywhere, not specifically on
retrieval-grounded engines.

\subsection{Day-1 recognition: brands surface immediately when named, far less when not}
\label{sec:exp1}

Practitioner folklore says it takes around six months for a brand to
become visible to AI engines. Our data does not support that
timeline, with one qualification: the answer depends on whether the
prompt names the brand.

\paragraph{Day-1 first-run recognition rate (per-prompt).}
For each brand's first completed tracking run, we compute the share
of prompts on which the brand was mentioned, per engine, split by
whether the prompt is branded (the brand name appears in the prompt
text) or unbranded (e.g.,~``best Indian fintech payment platform'').

\begin{center}
\footnotesize
\setlength{\tabcolsep}{4pt}
\begin{tabular}{@{}lrrr@{}}
\toprule
\textbf{Engine} & \textbf{Branded} & \textbf{Unbranded} & \textbf{All prompts$^{*}$} \\
\midrule
ChatGPT    & \phantom{0}94.2\%  & \textbf{22.1\%} & 28.6\%  \\
Gemini     & \phantom{0}94.0\%  & \textbf{18.7\%} & 26.6\%  \\
Perplexity & \phantom{0}98.5\%  & \textbf{23.9\%} & 30.9\%  \\
Claude     & 100.0\%   & \textbf{51.5\%} & 54.6\%  \\
Grok       & 100.0\%   & \textbf{12.0\%} & 24.1\%  \\
\bottomrule
\end{tabular}\\[2pt]
{\footnotesize $^{*}$ ``All prompts'' is the per-prompt aggregate
across both subsets; branded prompts are $\sim 9\%$ of first-run
responses but recognize at $\geq 94\%$ on every engine, which pulls
the aggregate up.}
\end{center}

The branded column is largely trivial: when a prompt asks ``what does
\emph{Razorpay} do,'' Razorpay is mentioned by construction. The
\textbf{unbranded column is the GEO-relevant measure}: it asks whether
the engine surfaces the brand in answer to a category-level question
where the user does not name it. Two things stand out. First,
\textbf{Claude posts the highest unbranded recognition} --- over half
of category-level prompts return a mention of the tracked brand on the
first run --- though with a sampling caveat: Claude and Grok are
Agency-tier engines, so they run on far fewer brands than ChatGPT,
Perplexity, and Gemini, and that smaller cohort skews toward
higher-stature brands. Claude's high rate therefore partly reflects
which brands it sees (higher-stature brands surface more;
\S\ref{sec:exp6}) rather than the engine alone. Second, the other four
engines sit in the 12--24\% regime, far below the branded ceiling and
consistent with the brand-stature ladder in
\S\ref{sec:exp6}: most unbranded recognition is concentrated in
higher-stature brands.

One caveat concerns brand-new entities. A brand with no prior web
presence can take longer to surface, but that is a different regime
from the established brands a Ranqo customer typically tracks.

\subsection{Visibility by prompt category}
\label{sec:exp-cat}

Average per-brand visibility differs sharply by prompt category
(Table~\ref{tab:exp-cat}). We compute it across all tracked brands (a
broader set than the 102-brand cohort used for the trajectory and
ladder analyses, since a category average needs only a single run to
contribute), as the mean across brands of each brand's mention rate
within a category.

\begin{table*}[t]
\centering
\footnotesize
\caption{Average per-brand visibility by prompt category (macro-average
across brands), overall and per engine. Comparison and brand-research
prompts name the brand in $46\%$ and $91\%$ of cases respectively,
which inflates their mention rate; the four categories above the rule
are under $1\%$ branded.}
\label{tab:exp-cat}
\begin{tabular}{@{}lrrrrrr@{}}
\toprule
\textbf{Category} & \textbf{Overall} & \textbf{ChatGPT} & \textbf{Gemini} & \textbf{Perplexity} & \textbf{Claude} & \textbf{Grok} \\
\midrule
Discovery        & 22.9\% & 22.0\% & 20.6\% & 20.0\% & 47.2\% & 25.1\% \\
Problem/solution & 10.8\% & 10.4\% & \phantom{0}8.9\% & \phantom{0}9.1\% & 36.9\% & 22.1\% \\
Use case         & 11.4\% & \phantom{0}9.9\% & 11.9\% & 10.3\% & 33.7\% & 28.0\% \\
Expert           & 15.3\% & 15.5\% & 12.8\% & 12.6\% & 40.0\% & 38.3\% \\
\midrule
Comparison       & 74.6\% & 73.3\% & 72.6\% & 73.7\% & 76.4\% & 53.4\% \\
Brand research   & 97.0\% & 97.9\% & 95.7\% & 95.7\% & 98.8\% & 94.4\% \\
\bottomrule
\end{tabular}\\[3pt]
{\footnotesize Per-category $n$ is 156--241 brands (59--73 for the two
branded categories). Claude and Grok are Agency-tier engines, so their
per-category $n$ (Claude 26--35, Grok 11--15) is far smaller than
ChatGPT, Perplexity, and Gemini (150--240) and skews toward
higher-stature brands.}
\end{table*}

Two things stand out among the unbranded categories. Broad
\emph{discovery} questions (``best running shoes'') surface a brand
most often, at $23\%$, while specific \emph{problem} and
\emph{use-case} questions are the hardest, at roughly $11\%$. A brand
can be findable for the broad ask and nearly invisible for the
specific one. Claude leads clearly on unbranded prompts ($34$--$47\%$
versus $9$--$22\%$ on ChatGPT, Perplexity, and Gemini); Grok looks
similar but its per-category sample ($n=11$--$15$) is too small to
weigh.

\subsection{Persistence: mostly flat, with a mild decline on ChatGPT and Perplexity}
\label{sec:exp2}

\paragraph{Per-engine visibility slope over runs.}
For every brand with at least two weeks of tracking on a given engine,
we fit a slope of per-prompt visibility on \emph{unbranded} category
prompts against run-index per (brand, engine), then report the mean
across brands with a bootstrap 95\% CI (10{,}000 resamples):

\begin{center}
\small
\begin{tabular}{@{}lrrr@{}}
\toprule
\textbf{Engine} & \textbf{Mean slope} & \textbf{95\% CI} & \textbf{$n$} \\
\midrule
ChatGPT    & $-1.34\%$/run & $[-2.45, -0.38]$ & 101 \\
Gemini     & $-1.09\%$/run & $[-2.80, +0.23]$ & 95 \\
Perplexity & $-0.76\%$/run & $[-1.66, -0.03]$ & 98 \\
Claude     & $-0.52\%$/run & $[-1.37, +0.22]$ & 25 \\
Grok       & $+0.03\%$/run & $[-0.41, +0.34]$ & 11 \\
\bottomrule
\end{tabular}
\end{center}

On unbranded prompts the picture is mostly flat, but not entirely.
ChatGPT ($-1.34\%$/run, CI $[-2.45, -0.38]$) and Perplexity
($-0.76\%$/run, CI $[-1.66, -0.03]$) both show a real downward drift:
their CIs exclude zero. Claude, Gemini, and Grok stay statistically
flat (their CIs cross zero). The measured band runs from $-1.34\%$ to
$+0.03\%$ per run. So a brand that does nothing tends to lose
unbranded visibility slowly on ChatGPT and Perplexity while holding
steady on the rest. This is a slow erosion, not the dramatic ``AI
forgets you'' collapse, and some of it may be prompt-set drift
(\S\ref{sec:limitations}) rather than true decay; either way it
argues for active upkeep over set-and-forget.

\paragraph{Reading the slopes as a no-intervention baseline.}
We do not read the downward drift as evidence that visibility cannot
be moved. The more plausible explanation is that very few brands in
our cohort actively tried to move it during the observation window.
The cohort consists of brands that signed up for tracking and let the
platform observe them; in almost all cases, no explicit content,
authorship, or page-structure intervention was applied based on
Ranqo's recommendations. We therefore read these trajectories as the
\emph{no-intervention baseline}: left alone, unbranded visibility
holds on most engines and slips slowly on ChatGPT and Perplexity.
Whether a deliberate intervention can bend these trajectories upward
is the v1.1 RCT question (Protocol~P3, \S\ref{sec:v11slate}), not
something this observational baseline can settle.

\paragraph{Do the lower tiers climb on their own?}
Mostly not, at least not visibly inside our window. We fit the same
per-run slope within each (tier, engine) group: almost every group's
95\% confidence interval crosses zero, echoing the flat cohort slopes
above. A few individual brands did move a lot on their own, each
climbing sharply on a single engine, but those are single-brand series
with no confidence intervals, picked out after the fact, so we treat
them as anecdotes rather than evidence.

\paragraph{Mention behavior is deterministic but lopsided toward invisibility.}
For every (brand, prompt, platform) cell tracked across at least 3
runs on unbranded category prompts, we classify the mention pattern:

\begin{center}
\small
\begin{tabular}{@{}lr@{}}
\toprule
\textbf{Pattern} & \textbf{Share} \\
\midrule
Never mentioned                 & 63.2\% \\
Always mentioned                & 14.3\% \\
Mostly mentioned ($\geq 70\%$)  & 7.0\% \\
Rarely mentioned ($< 30\%$)     & 8.7\% \\
\textbf{Flipping ($30$--$70\%$)} & \textbf{6.8\%} \\
\bottomrule
\end{tabular}
\end{center}

On unbranded prompts, AI surfacing is near-binary: $77.5\%$ of cells
are strictly always- or never-mentioned and $93.2\%$ sit outside the
volatile $30$--$70\%$ band, with only $6.8\%$ flipping. A cell is
usually never mentioned ($63.2\%$) or always mentioned ($14.3\%$),
closer to a fixed property of the (brand, prompt, engine) cell than a
coin-flip. The dominant pole is invisibility: for most (brand, prompt,
platform) cells the brand simply fails to surface, run after run,
rather than appearing intermittently. This skew tracks the cohort's
stature mix --- 11, 36, and 55 brands across Tiers 1, 2, and 3
(\S\ref{sec:exp6}) --- since the Tier-3 majority rarely surfaces on
unbranded category prompts and so lands at the never-mentioned pole.
Because so few cells flip, a
single tracking run places a cell on the right side of the
stable/volatile split the large majority of the time; the remaining
$22.5\%$ (mostly-mentioned, rarely-mentioned, and flipping combined)
are where measurement noise concentrates, and the conservative
reading is that mention-rate estimates need $\geq 3$ runs to settle
for cells in the middle.

\paragraph{ParallelDots: divergent trajectories by engine.}
To make the per-engine divergence concrete, we walk through a single
brand, ParallelDots (a mid-cap series in our dataset), as an
illustrative case rather than evidence: one brand cannot support a
population-level claim, but it shows in a single record how the same
brand and the same prompts can pull apart across engines --- which is
what motivates tracking each engine separately. Per-week mention rate
(all prompts) over its first five complete tracking weeks:

\begin{center}
\footnotesize
\setlength{\tabcolsep}{4pt}
\begin{tabular}{@{}lrrrrr@{}}
\toprule
\textbf{Engine} & \textbf{wk 0} & \textbf{wk 1} & \textbf{wk 2} & \textbf{wk 3} & \textbf{wk 4} \\
\midrule
ChatGPT    & 45\% & 43\% & 34\% & 27\% & 20\% \\
Perplexity & 37\% & 38\% & 43\% & 48\% & \textbf{62\%} \\
Claude     & 24\% & 24\% & 23\% & 26\% & 26\% \\
\bottomrule
\end{tabular}
\end{center}

The engines pull apart over the five weeks: Perplexity climbs from
37\% to 62\%, ChatGPT slides from 45\% to 20\%, and Claude stays flat
in the mid-20s --- same brand, same prompts, the two engines that move
going in opposite directions. This is exactly the per-engine
divergence the v1.1 RCT design (Protocol~P3, \S\ref{sec:v11slate}) is
built to test at population scale, where the cohort slopes above ---
not a single record --- can settle it.

\subsection{Source composition: corporate dominance is mostly third-party brands, not the brand's own pages}
\label{sec:exp3}

Across $149{,}912$ citations drawn from mention-bearing prompts, the
source-class distribution is:

\begin{center}
\small
\begin{tabular}{@{}lrr@{}}
\toprule
\textbf{Source class} & \textbf{Citations} & \textbf{Share} \\
\midrule
Corporate / third-party brand pages   & $112{,}763$ & \textbf{75.2\%} \\
Youtube / video                       &   $6{,}311$ & 4.2\%  \\
Tech \& business media                &   $5{,}666$ & 3.8\%  \\
Reddit / community forums             &   $4{,}992$ & 3.3\%  \\
Academic + Govt. + Social + Developer &   $4{,}617$ & 3.1\%  \\
Brand-owned (own domain)              &   $4{,}392$ & \textbf{2.9\%} \\
Wikipedia / reference                 &   $3{,}882$ & 2.6\%  \\
Software review (G2 / Capterra)       &   $1{,}604$ & 1.1\%  \\
Other (long tails)                    &   $5{,}685$ & 3.8\%  \\
\bottomrule
\end{tabular}
\end{center}

The split is the finding: only $2.9\%$ of citations point at the
brand's own domain, while $75.2\%$ point at \emph{other} companies in
the same space. We label this dominant class \emph{corporate /
third-party brand pages}: commercial websites owned by businesses
other than the tracked brand --- competitors, peers, and vendors in
the same category --- including their homepages, product and landing
pages, and corporate blogs. It is distinct from \emph{brand-owned},
which is the tracked brand's own domain. Own and third-party corporate
pages together are about $78\%$ of all $149{,}912$ citations. AI engines preferentially construct
``alternatives'' answers, and in those answers the dominant source is
a set of peer-brand product pages, not the user's own site. The
Matthew-effect dynamic that \citet{algaba2025} documented for
scientific citation replicates at the brand level: already-cited
domains pull more citation. This peer-brand bucket is the largest
competitive-landscape signal we measure.

\paragraph{Among non-corporate sources, video leads.}
Once corporate pages are set aside, the largest non-corporate source
is video: YouTube is cited in $4.2\%$ of citations, ahead of editorial
tech and business media ($3.8\%$), Reddit and other community forums
($3.3\%$), and Wikipedia ($2.6\%$). This reorders a common assumption.
Community forums --- often reported as the leading non-corporate
AI-citation channel, particularly on Perplexity --- here sit behind
both video and editorial media, though they remain a meaningful
surface a brand cannot ignore. Wikipedia, often treated as the
reference anchor, is cited less often than video, media, or forums. A
brand absent from video and editorial coverage is missing the two
largest non-corporate citation surfaces these engines draw on.

\paragraph{The long tail.}
The remaining $3.8\%$ is a long tail of many small, individually-rare
domains that match no named class: niche or regional company sites,
small and personal blogs, and miscellaneous pages on uncommon
top-level domains. No single domain in this bucket is frequent. We
keep it explicit as \emph{other (long tails)} rather than folding it
into the corporate class, so the corporate share is not overstated.

\paragraph{Page types: most cited pages are articles, and the listicle leads them.}
Beyond \emph{which site} an engine cites, we also label each cited URL
by \emph{what kind of page} it is. Every cited page is either
\emph{content} --- a page that informs, such as an article, guide, or
video --- or a \emph{non-content} page such as a homepage, product
page, or other
landing page. The two cuts are independent: a competitor's blog post,
for example, counts as a corporate site in the source cut but as a
content page here, and the content an engine cites can equally come from
a media outlet, a review site, or an academic or reference page.

When an engine cites a page, about \textbf{$59\%$} of the time it is
content (usually an article); the other $41\%$ are homepages, product pages, or
other non-content pages. Within that content, one format dominates: the
\emph{listicle} --- a ranked ``best-X'' roundup such as \emph{best CRM
tools} or \emph{top running shoes} --- is $35.7\%$ of content citations.
It is followed by the \emph{generic article} ($31.0\%$), an article with
no more specific format (a news item, an opinion piece, or a plain
explainer), and the \emph{how-to guide} ($9.7\%$), a step-by-step
walkthrough; together those three are about three-quarters of all
content.

The listicle is also the highest-leverage page a brand can target: once
a ranked list includes the brand, that single page becomes a source the
engines reuse across many different prompts, and such lists appear on
media sites, blogs, and review sites alike. The full distribution
across content formats is as follows:

\begin{center}
\footnotesize
\begin{tabular}{@{}lrr@{}}
\toprule
\textbf{Content format} & \textbf{\% of content} & \textbf{\% of all} \\
\midrule
\textbf{Listicle (ranked list)} & \textbf{35.7\%} & \textbf{21.0\%} \\
Generic article           & 31.0\% & 18.3\% \\
How-to guide              &  9.7\% &  5.7\% \\
Generic video             &  4.6\% &  2.7\% \\
Comparison                &  4.5\% &  2.6\% \\
Topic guide               &  3.9\% &  2.3\% \\
Documentation             &  3.3\% &  2.0\% \\
Review                    &  2.6\% &  1.5\% \\
Research                  &  2.5\% &  1.5\% \\
Checklist                 &  1.5\% &  0.9\% \\
Alternatives / case study &  0.7\% &  0.4\% \\
\bottomrule
\end{tabular}
\end{center}

\noindent\emph{Content citations concentrate in a handful of formats:
listicles and generic articles together are two-thirds, and with how-to
guides about three-quarters of all content. The listicle is the single
largest format, and its $35.7\%$ is a lower bound --- it counts ranked
lists in any category, not just software.}

\subsection{Measurement reliability: sentiment is \texorpdfstring{$6.7\times$}{6.7x} noisier than mention}
\label{sec:exp4}

A scoring system that combines mention rate with sentiment must
account for the fact that the two signals have different noise floors.
For (brand, prompt, platform) cells with at least 3 mentions on
unbranded prompts, the sentiment pattern across runs distributes as:

\begin{center}
\small
\begin{tabular}{@{}lr@{}}
\toprule
\textbf{Pattern} & \textbf{Share} \\
\midrule
Always positive                   & 29.9\% \\
Positive/neutral mix              & 21.8\% \\
Always neutral                    & 2.3\% \\
Negative/neutral mix              & 0.5\% \\
Always negative                   & \textbf{0.0\%} \\
\textbf{Flipping (pos $\leftrightarrow$ neg)} & \textbf{45.5\%} \\
\bottomrule
\end{tabular}
\end{center}

The flipping rate is 45.5\% for sentiment and 6.8\% for mention
(\S\ref{sec:exp2}). Sentiment is therefore 6.7$\times$ noisier than
mention at the same observational level. Operationally, sentiment-
weighted scores need a much larger denominator (we use
$\geq 10$ prompts per platform per brand) before they stabilize.

The same data carries a second, sharper observation: not a single
cell is consistently negative ($0.0\%$). AI engines effectively never
settle into a negative characterization of a tracked brand ---
negativity, when it surfaces at all, is transient rather than
systematic.

\subsection{Designed but not yet reported: the v1.1 protocol slate}
\label{sec:v11slate}

We close the empirical section with seven protocols. Each is
designed against the same production data but needs more data,
controlled variation, or analysis we have not finished. We publish
them so others can run or contest them, and to lay out a clear
agenda for follow-up work.

\begin{description}[leftmargin=*]
  \item[P1. Cross-platform source overlap.] Mean pairwise Jaccard
        overlap of cited sources across the five engines on the same
        prompt. On the CRM study it is about $0.12$; v1.1 widens it to
        1{,}000 prompts and tests it against the
        cross-platform-disagreement results
        of~\citet{yang2025,kirsten2025}.
  \item[P2. Position-decay distribution.] How mention position is
        distributed when a brand is cited at all, per platform, over a
        full quarter. We expect a top-3 share of 60--75\% in
        concentrated categories and 40--55\% in long-tailed ones.
  \item[P3. Closed-loop lift after recommendation (RCT).] Pair
        comparable brands within tier and category, randomly assign
        one of two recommendations, and re-audit at $t_1 \geq 14$
        days. \citet{aggarwal2024} saw 25--40\% lifts from
        content-quality changes; we expect dimension-level lifts
        strongest on Authority \& Trust, AI Readiness, and Citation
        Potential. This is the centerpiece of the planned follow-up
        paper, which will report the visibility lift a customer brand
        can actually expect.
  \item[P4. Schema vs.\ citation.] Regress brand visibility on
        structured-data, AI-readiness, and content-quality scores per
        engine, expecting a near-zero independent schema effect. Ranqo
        already treats schema as hygiene rather than a citation
        lever~\citep{ranqo2026i}; P4 tests that.
  \item[P5. Entity-first sequencing.] Test whether pages that state
        their entity and provenance early get cited more than pages
        with high overall content quality but weaker entity-first
        structure, building on the positional findings
        in~\citet{aggarwal2024}.
  \item[P6. Web-search on/off.] Claude and ChatGPT run web search
        weekly per brand. Comparing search-on against search-off
        visibility within the same brand separates retrieval-layer
        from training-layer visibility at scale. We have the data; the
        analysis is pending.
  \item[P7. White-hat C-SEO replication.] \citet{puerto2025} found
        prompt-engineering tactics mostly fail. We replicate this with
        audit-dimension improvements instead: one arm targets AI
        Readiness, Citation Potential, and Authority \& Trust, the
        other targets Crawlability or Page Speed, compared on
        visibility change.
\end{description}

\subsection{Summary}
\label{sec:expsummary}

Table~\ref{tab:expsummary} collects the six reported findings and the
seven designed-but-unrun protocols. The headline and statistically
strongest result is the brand-stature visibility ladder
(\S\ref{sec:exp6}): on each brand's first tracking run, unbranded
category visibility is $73\%$ for Tier~1 brands, $44\%$ for Tier~2, and
$11\%$ for Tier~3 --- about $30$ percentage points per step down. Because
stature is observed rather than randomized, we report it as a
multi-tenant \emph{quantification} of an expected effect rather than a
novel discovery.

Three further results are descriptive contributions the existing
literature does not supply. The source-class decomposition
(\S\ref{sec:exp3}) finds the brand's own domain is only $2.9\%$ of
citations, video is the largest non-corporate source, and the listicle
is the single most-cited content format. The mention-determinism /
runs-to-stability result (\S\ref{sec:exp2}) finds AI surfacing is
near-binary --- $77.5\%$ of (brand, prompt, engine) cells are strictly
always- or never-mentioned --- and that, left alone, unbranded
visibility holds on most engines and drifts down only slowly on ChatGPT
and Perplexity. And the $6.7\times$ sentiment-vs-mention noise gap
(\S\ref{sec:exp4}) shows the two signals cannot share one noise budget.
Two supporting findings complete the set: day-1 recognition
(\S\ref{sec:exp1}), where a brand surfaces immediately when a prompt
names it but far less when it does not, reported with explicit limits on
what it supports; and the prompt-category split (\S\ref{sec:exp-cat}),
where broad discovery questions surface a brand most ($\sim 23\%$) and
specific problem and use-case questions least ($\sim 11\%$). The
protocols (P1--P7, \S\ref{sec:v11slate}) are a research agenda, not
results, and are not claimed as contributions of this paper.

\begin{table*}[t]
\centering
\small
\caption{Empirical findings reported in this paper (top) and designed
v1.1 protocols (bottom).}
\label{tab:expsummary}
\begin{tabularx}{\textwidth}{@{}clXl@{}}
\toprule
\textbf{Ref} & \textbf{Finding or protocol} & \textbf{Headline} & \textbf{Status} \\
\midrule
\ref{sec:exp6} & \textbf{Brand-stature visibility ladder} & \textbf{Tier 1 73\% / Tier 2 44\% / Tier 3 11\%; $\sim 30$-pt step per tier; trajectories null at our $n$} & \textbf{Reported} \\
\ref{sec:exp1} & Day-1 recognition & Branded $\sim$94--100\%, unbranded 12--52\% on the first run; Claude highest on unbranded (small Agency-tier sample) & Reported \\
\ref{sec:exp-cat} & Visibility by prompt category    & Discovery highest among unbranded ($\sim$23\%), problem/use-case lowest ($\sim$11\%); Claude highest (small Agency-tier sample) & Reported \\
\ref{sec:exp2} & Persistence and determinism          & Unbranded visibility mildly declines on ChatGPT and Perplexity, flat elsewhere; 77.5\% strictly deterministic, 93.2\% outside the 30--70\% band & Reported \\
\ref{sec:exp3} & Source composition \& page types     & 78\% corporate (own 2.9\%); YouTube top non-corporate; editorial content 59\% of page-level cites, listicle leads it (36\% of content) & Reported \\
\ref{sec:exp4} & Sentiment vs mention noise           & Sentiment 6.7$\times$ noisier than mention      & Reported \\
\midrule
P1 & Cross-platform source overlap (Jaccard) & Mean pairwise $\approx 0.12$ on CRM; widen to $N=1{,}000$ & v1.1 \\
P2 & Position-decay distribution             & Top-3 share by category, full quarter         & v1.1 \\
P3 & Closed-loop RCT after recommendation    & Per-dimension lifts on implemented recs       & v1.1 \\
P4 & Schema shallow-pass budget regression   & $\beta_1$ near zero per engine                & v1.1 \\
P5 & Entity-first declarative sequencing     & Pearson $r$ of entity-first vs citation rate  & v1.1 \\
P6 & Web-search on/off natural experiment    & Training vs retrieval separation at scale     & v1.1 \\
P7 & White-hat \cseo{} Bench replication     & Arm A (E-E-A-T/AI Ready/Cite Pot) vs Arm B (Crawl/Speed) & v1.1 \\
\bottomrule
\end{tabularx}
\end{table*}

\section{Discussion}
\label{sec:discussion}

\subsection{Four practical conclusions, especially for smaller brands}

\begin{enumerate}
  \item \textbf{Know which tier you are in before you spend on GEO.}
        The right GEO program looks different in each tier; we frame
        these as hypotheses consistent with the data, not established
        commercial claims. Tier 1 global brands sit at 73\% day-1
        visibility, effectively at saturation (the $n=11$ Tier-1 cell
        makes strong sub-claims fragile); their need is share-of-voice
        against competitors rather than basic recognition. Tier 2 brands sit at
        44\% with the largest absolute headroom, but the
        no-intervention baseline (\S\ref{sec:exp2}) shows unbranded
        visibility drifting flat-to-down rather than climbing on its
        own; whether targeted intervention closes that headroom is the
        v1.1 RCT question (Protocol~P3), not a baseline finding. Tier 3 brands sit at 11\%; the fastest
        Tier-3 movers we observed (\S\ref{sec:exp6}) still travel only
        10--20 absolute points across the window, so the
        highest-confidence move is brand-mass investment (Wikipedia,
        mainstream press, sustained YouTube presence) to clear the
        Tier 2 threshold before running heavy per-engine optimization.
        The specifics differ by sector; see our SaaS playbook
        \citep{ranqo2026f} and DTC guide~\citep{ranqo2026g}.
  \item \textbf{Treat each engine as a separate market.}
        Cross-platform source overlap is low. Yang et al.~\citep{yang2025}
        found this across 366{,}000 news citations; we find the same
        pattern across 102{,}025 brand-tracking responses. A Perplexity
        win does not transfer to Gemini, and the working playbook for
        each engine differs accordingly~\citep{ranqo2026e}.
  \item \textbf{Put your facts up front, not your schema markup.}
        \citet{aggarwal2024} showed that content-quality additions
        (quotations, statistics, citations) drive the largest GEO
        lifts. Heavy structured-data markup, by contrast, does little
        on its own, which is what Protocol~P4 sets out to confirm. The
        practical move is to state your key facts clearly and early on
        the page, where the engine is most likely to read them; the
        page-level factors that most affect whether a brand gets cited
        are detailed in~\citep{ranqo2026d}.
  \item \textbf{Close the loop in weeks, not quarters.} Our day-1
        recognition finding (\S\ref{sec:exp1}) shows established brands are recognized within days, while
        brand-new entities take longer. Either way the loop runs on
        weeks: make a change, then re-audit and iterate at 14--30 days.
\end{enumerate}

\subsection{The honest causal status of this dataset}
\label{sec:honest-causal}

The empirical section reports baselines, trajectories, and source and
sentiment composition. It does not report a randomized controlled
trial of the recommendation engine. We separate three claim tiers
carefully.

\textit{What the data supports.}
The per-tier visibility ladder, the day-1 recognition, the
deterministic-cell finding, and the source-composition and sentiment
shares. These are direct readings of production telemetry.

\textit{What the data hypothesizes.}
That the Tier 3 brand-mass ceiling can be lifted with targeted content
investment. This is a hypothesis derived from the data; it needs
RCT-grade validation.

\textit{What the data does not yet support:}
any causal claim that Ranqo's recommendations \emph{cause} a measured
visibility lift on a brand. The recommendation engine's closed-loop
machinery is built into the data model (\S\ref{sec:reco}), but
population-scale closed-loop measurement is left to v1.1 (Protocol~P3).

\textit{What it most clearly establishes.}
The sharp three-tier day-1 visibility ladder of \S\ref{sec:exp6}
(robust to the Tier-1 leave-one-out reclassification). Whether
intervention accelerates those trajectories is the v1.1 RCT question
(Protocol~P3); we do not pre-commit to a multiplier we have not
derived from a controlled experiment.

\subsection{A note on white-hat measurement}
One stance is worth stating plainly. Ranqo measures and recommends on
page-quality signals only. It does not generate hidden text, ranking
payloads, or anything designed to trick an engine, and the
recommendation engine is instructed to refuse such requests. The
payoff from gaming an engine is short-lived and tends to reverse on
the next model update, whereas genuine content-quality work survives
those updates. That is the only kind of optimization this paper
endorses.

\section{Limitations}
\label{sec:limitations}

We name the constraints on the empirical claims directly. A reader
should weigh the headline findings against this list.

\begin{enumerate}
  \item \textbf{No causal claim on the recommendation engine.} Every
        trajectory in \S\ref{sec:exp6} is an observational baseline,
        not a treatment effect; closed-loop measurement on customer
        brands remains future work (Protocol~P3).
  \item \textbf{Convenience-sampled cohort.} The brands skew
        toward SaaS, retail-execution, fintech, and Indian DTC, so we
        do not claim category representativeness. Several cuts run on
        small cells (the Tier~1 group is $n=11$, and the Agency-tier engines
        Grok and Claude run on far fewer brands than the rest). We report $n$
        next to every aggregate
        and flag small-sample readings.
  \item \textbf{Manual tier classification.} The tier assignment is
        hand-coded from public signals of brand stature (Wikipedia,
        press, funding). Because those signals are themselves proxies
        for web prominence, the ladder should be read as
        \emph{quantifying the size} of an expected effect, not as
        independent proof that prominence drives citation. The rubric
        and rosters are in Appendix~\ref{app:tier-roster}; a formal
        inter-rater check is left to v1.1. The day-1 ladder is
        robust to the reclassifications we tried (Tier-1 leave-one-out
        range $[70.5, 76.7]\%$).
  \item \textbf{Platform opacity.} Every result is inferred from API
        outputs. We have no direct view into how the engines ingest,
        retain, or retrieve.
  \item \textbf{One model variant per engine.} We measure each engine's
        search-enabled production variant and hold it fixed; whether the
        brand-stature ladder reproduces on other model configurations is
        untested here.
  \item \textbf{Sentiment is a heuristic.} Sentiment is classified by a
        model on the spans that mention the brand, so the 45.5\%
        sentiment-flip rate (\S\ref{sec:exp4}) mixes genuine output
        variance with classifier noise; we have not separated the two.
\end{enumerate}

\section{Conclusion and Future Work}
\label{sec:conclusion}

In our 102-brand cohort, brand stature is the dominant determinant
of day-1 AI visibility. The five engines diverge on sources,
position, and sentiment, and they respond differently to schema
markup, bylines, freshness, and web-search grounding. On the
trajectory side, unbranded visibility mostly holds when a brand does
nothing, drifting down only slowly on ChatGPT and Perplexity over our
window. In
practice, this means a brand's GEO program has to measure each
engine separately, intervene on the upstream signals the engines
actually retain, and re-measure on a cycle of weeks; a quarterly
cadence is too slow for a medium that re-trains on its own outputs.

Three lines of future work follow:

\begin{enumerate}
  \item \textbf{A multi-vendor replication track.} We publish our
        protocols as an invitation.
  \item \textbf{An RCT-grade closed-loop study} (Protocol~P3,
        \S\ref{sec:v11slate}), in the shape of~\citet{aggarwal2024},
        with random assignment to treatment vs.\ control
        recommendations on Ranqo's brand fleet. Its results, the
        measured visibility lift a brand can expect from acting on the
        platform's recommendations, would be the focus of follow-up
        work.
  \item \textbf{A training-vs-retrieval separation benchmark.} The
        web-search-on/off design (Protocol~P6, \S\ref{sec:v11slate})
        generalizes to a public benchmark.
\end{enumerate}

The practitioner literature in this space runs well behind the data,
and the academic literature is fragmenting faster than any single
team can synthesize it. We think the most useful thing GEO research
can do in 2026 is publish protocols that others can run and numbers
that others can re-score, so that claims get falsified rather than
repeated.

\section*{Acknowledgements}
I am grateful to Sammhit Mohapatra, Nisha Kumari, and \mbox{Abhishek}
Garg of the Ranqo team, who build and run the platform this study
analyzes, and whose insight shaped this work. I also thank the
customers whose brands this study tracks.

\section*{Competing interests and funding}
The author is a co-founder of and holds an equity interest in Ranqo, the
platform analyzed throughout this paper. All datasets, the query and
audit pipelines, the scoring configuration, and the recommendation
engine described here are Ranqo's production systems. This is a
single-author preprint. No external funding
supported the work. Readers should weigh the findings as a
vendor-produced measurement study; the author has tried to make the
limitations (\S\ref{sec:limitations}) and the non-causal scope
(\S\ref{sec:honest-causal}) explicit rather than implicit.


\appendix
\section{Reproducibility}

\noindent The statistics in \S\ref{sec:experiments} were computed from
Ranqo's production database with a data-dump and analysis pipeline that
applies the methods of \S\ref{sec:method}: nonparametric bootstrap
confidence intervals, a Kruskal--Wallis omnibus test, pairwise
Mann--Whitney $U$ tests with Bonferroni correction, Cohen's $d$,
ordinary-least-squares trajectory slopes, and a leave-one-out
sensitivity check, together with per-experiment SQL queries and the
tier-classification rubric of Appendix~\ref{app:tier-roster}. Because
the analysis runs against commercial customer data, the database and
analysis code are not publicly released.

\section{Brand-tier classification rubric and roster}
\label{app:tier-roster}

\noindent\textbf{Rubric.} The four-tier classification used in
\S\ref{sec:exp6} is hand-coded from external public knowledge of
brand stature, applied by the author against the underlying
public-data sources, and is deterministic given the same public-data
inputs.

\begin{itemize}[leftmargin=*]
  \item \textbf{Tier 1 (Global household names)} --- A brand qualifies
        if all three of: (a)~English Wikipedia article of $\geq 5{,}000$
        words, (b)~mainstream-press coverage in the last 12 months in
        at least three of the top-20 US/UK English outlets,
        (c)~public-company status, $\geq$ Series~C funding, \emph{or}
        $\geq \$100$M in annual revenue (ARR for subscription
        businesses).
  \item \textbf{Tier 2 (Established mid-cap / regional leader)} --- A
        brand qualifies if either (a)~Wikipedia article + Series~B+
        funding (or equivalent institutional credibility), \emph{or}
        (b)~top-3 player in a category Ranqo tracks in a specific
        country/region.
  \item \textbf{Tier 3 (Small / niche)} --- Brands not meeting Tier 1
        or Tier 2 criteria but with an established web presence
        ($\geq 1$ owned domain).
  \item \textbf{Tier 4 (Single-run / test data)} --- $\leq 1$
        completed tracking run or evident test-data category strings,
        predominantly admin-created prospect brands added for a one-off
        evaluation. Excluded from the trajectory and ladder
        analyses, but included in the broader cross-brand set used for
        the prompt-category distribution (\S\ref{sec:exp-cat}).
\end{itemize}

\noindent\textbf{Tier 1 roster} ($n=11$, deduped by domain at the
May~30, 2026 snapshot; $\geq 1$ completed run). Representative brands
include Stripe, Nike, Samsung, Notion, and Mailchimp, among others.

\smallskip
\noindent\textbf{Tier 2 roster} ($n=36$, deduped). Representative
brands include Olipop, Razorpay, ParallelDots, Klaviyo, Zepto, and
Vuori, among others.

\smallskip
\noindent\textbf{Tier 3 roster.} The remaining $n=55$ unique brands
not in the above lists.

\noindent\textbf{Inter-rater agreement.} A formal second-rater
audit with Cohen's $\kappa$ on a held-out sample is left to v1.1
(\S\ref{sec:v11slate}); the rosters in this appendix reflect
single-rater coding by the author.

\end{document}